\documentclass{acmart}
\setcopyright{none}
\makeatletter
\def\@authorsaddresses{}
\makeatother
\renewcommand\footnotetextcopyrightpermission[1]{}

\usepackage{enumitem}
\usepackage{booktabs}
\usepackage{multirow}
\usepackage{rotating}
\usepackage{hyperref}
\usepackage{changepage}

\usepackage{tabularx}
\usepackage{booktabs}
\usepackage{array}
\usepackage[table]{xcolor}
\usepackage{enumitem}
\usepackage{ragged2e}

\definecolor{Mint}{HTML}{E9F7F0}
\definecolor{Butter}{HTML}{FFF6DA}
\definecolor{Lav}{HTML}{EEEAFB}
\definecolor{Sage}{HTML}{E5F3E0}  
\definecolor{LightGray}{HTML}{F2F2F2} 
\definecolor{SoftBlue}{HTML}{D7E6F5}  
\definecolor{SoftGreen}{HTML}{D9F0E5}

\setlist[itemize]{leftmargin=*, itemsep=2pt, topsep=2pt, parsep=0pt}

\AtBeginDocument{%
  }

\begin{document}


\title{Practitioners' Perspectives on Designing Data Visualizations for the General Public}

\author{Regina Schuster}
\affiliation{%
  \department{Faculty of Computer Science, Doctoral School Computer Science}
  \institution{University of Vienna}
  \city{Vienna}
  \country{Austria}
}
\email{regina.maria.veronika.schuster@univie.ac.at}

\author{Kathleen Gregory}
\affiliation{%
  \department{Centre for Science and Technology Studies}
  \institution{Leiden University}
  \city{Leiden}
  \country{Netherlands}
}
\email{k.m.gregory@cwts.leidenuniv.nl}

\author{Torsten Möller}
\affiliation{%
  \department{Faculty of Computer Science, Data Science @ Uni Vienna}
  \institution{University of Vienna}
  \city{Vienna}
  \country{Austria}
}
\email{torsten.moeller@univie.ac.at}

\author{Laura Koesten}
\affiliation{%
  \department{Department of Human-Computer Interaction}
  \institution{Mohamed bin Zayed University of Artificial Intelligence}
  \city{Abu Dhabi}
  \country{United Arab Emirates}
}
\affiliation{%
  \institution{AIT Austrian Institute of Technology}
  \city{Vienna}
  \country{Austria}
}
\affiliation{%
  \department{Faculty of Computer Science}
  \institution{University of Vienna}
  \city{Vienna}
  \country{Austria}
}
\email{laura.koesten@mbzuai.ac.ae}

\renewcommand{\shortauthors}{Schuster et al.}

\begin{abstract}
Public-facing data visualizations can play a vital role in making complex information clear and engaging, thereby encouraging informed public discourse and participation. However, existing work offers limited insight into how practitioners make design decisions based on their envisioned target audiences and across different media channels. To investigate this, we conducted semi-structured interviews with 21 professionals from journalistic settings, focusing on how they conceptualize their readers, translate these notions into design choices, and evaluate their work. We found that practitioners often rely on broad audience definitions, despite considering ``knowing their readers'' essential. Evaluation primarily relies on peer feedback or social metrics rather than user testing. From these accounts, we identify recurring strategies employed to reach general, often undefined publics. We discuss implications for audience-centered authoring tools, proposing features such as persona simulations and content-adaptive multi-format authoring, message-first rhetoric-aware workflows, and lightweight in-tool evaluation to better support the realities of public-facing design.
\end{abstract}




\begin{teaserfigure}
  \centering
  \includegraphics[width=\textwidth]{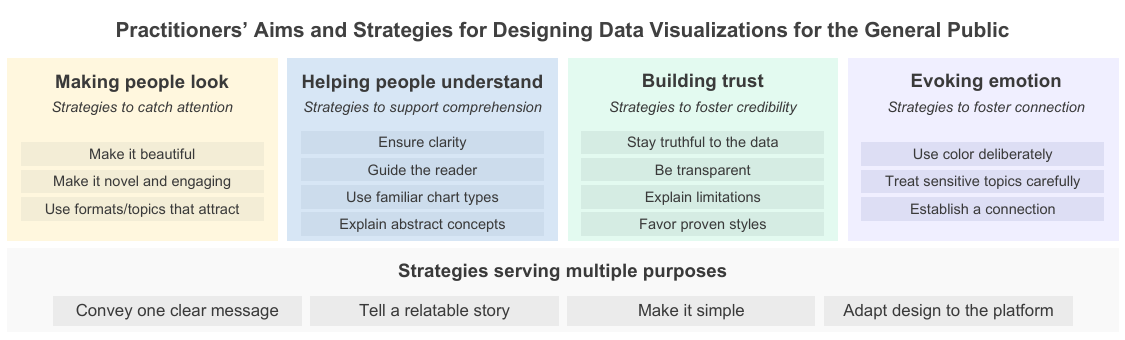}
  \caption{Four aims of visualization design that practitioners commonly discussed: making people look, helping people understand, building trust, and evoking emotion. For each aim, the strategies most frequently mentioned in the interviews are shown. Practitioners also described strategies serving multiple purposes across aims, and noted that overlaps and tensions arise between different aims. Strategies are further described in \autoref{tab:strategies}.}
  \Description{Four aims of visualization design and associated practitioner strategies. The first box describes making people look and includes strategies such as making visualizations visually appealing, using novel or engaging formats, and choosing attention-grabbing topics. The second box describes helping people understand and includes strategies such as ensuring clarity, guiding the reader, using familiar chart types, and explaining abstract concepts. The third box describes building trust and includes strategies such as staying truthful to the data, being transparent, explaining limitations, and using established visual styles. The fourth box describes evoking emotion and includes strategies such as deliberate use of color, careful treatment of sensitive topics, and establishing personal connection. A separate box lists cross-cutting strategies used across aims, including conveying one clear message, telling relatable stories, simplifying design, and adapting design to the platform.}
  \label{fig:teaser}
\end{teaserfigure}


\maketitle

\section{Motivation}
Public-facing data visualizations play a vital role in communicating data to the general public. Such visualizations, which are geared toward \textit{``everyday users''}~\citep{pousman_casual_2007}, have become staple features in news media~\citep{fu_more_2024,lopezosa_data_2023,engebretsen_data_2020}. This trend has only been magnified by the COVID-19 pandemic, as journalists worked to communicate complex data visually to the public through a variety of media channels~\citep{zhang_mapping_2021,yang_systematic_2022}. In response, a growing body of research has begun to investigate the production of (COVID-19) data visualizations for general audiences~\citep[e.g.,][]{zhang_visualization_2022,parsons_understanding_2021,gregory_datajourneys_2024}. These studies demonstrate that such visualizations result from situated design practices, where practitioners navigate data, visualization design guidelines, tools, constraints, and audience expectations.

In this paper, we investigate the design of such public-facing data visualizations, i.e., graphics that are intended for broad audiences in journalistic and media settings. Previous work in this area has often used the term ``casual information visualization,'' although definitions vary in emphasis~\citep[e.g.,][]{pousman_casual_2007,sprague_exploring_2012,trajkova_exploring_2020}. The framing as ``casual,'' however, may unintentionally suggest that such visualizations are inherently simple or less serious and hence potentially reinforce an unnecessary dichotomy. To avoid this, we use the term ``public-facing,''~\citep[e.g.,][]{bullock_exploring_2022, zhang_visualization_2023, strantz_using_2021, cabric_eleven_2024}. Importantly, the notion of ``a public'' is also far from monolithic, as there is no single general audience and attempts to design for ``everyone'' often exclude people with different levels of literacy, interest, or ability. We therefore use ``public-facing'' to refer not to a uniform user group, but to a context of engagement: non-work, discretionary settings in which people encounter data visualizations as part of everyday media use (such as reading news on a mobile device while on public transport). Accordingly, we define the scope of this study as data visualizations created for broad, heterogeneous audiences who may or may not have specific topic expertise or visualization literacy.

Given this heterogeneity, we expect that creators of public-facing data visualizations, particularly in journalistic settings, also rely on broad notions of their readership, as media cater to diverse publics. However, this seems at odds with the common principle in visualization research that effective visualization requires ``knowing your audience''~\citep{drucker_communicating_2018,lee_reaching_2020,bottinger_reflections_2020}. This raises questions about what strategies practitioners actually employ when designing for such broad publics.

Designers may turn to principles grounded in empirical visualization research or to established best practice guidelines. Several have condensed their experience into handbooks; the most famous, perhaps, are Tufte's works~\citep{tufte_visual_1983,tufte_envisioning_1990,tufte_explanations_2005}, but there is a long list of practitioner-oriented books~\citep[e.g.,][]{schwabish_better_2021,cairo_functional_2012,few_show_2012}. However, such guidelines and the tools built around them do not necessarily align with the realities of journalistic production. Understanding how design choices are made is crucial not only for bridging research and practice but also for society at large. Data visualizations influence how the public encounters and interprets issues such as climate change, health, or inequality, raising questions of fairness and access~\citep{dignazio_data_2023}. If data are meant to inform ``all'' citizens, visualizations must be approachable to diverse audiences, including visualization novices~\citep{burns_who_2023}.

This makes it essential to understand how audience needs are addressed in practice and where current support systems fall short. Previous research has called for future work to investigate what designers need to know about audiences to create understandable and attractive visual representations~\citep{burns_designing_2022} and how to adapt them for different media channels~\citep{schuster_being_2024,de_haan_when_2018}. Yet, little is known about how creators of public-facing data visualizations in journalism make design choices, including the concrete strategies, guiding principles, and professional constraints involved. Research also increasingly emphasizes the involvement of users in evaluating chart design~\citep{ridley_evaluating_2020}; however, it is unclear how journalistic visualizations are, or can be, evaluated by broadly defined target audiences, given the fast-paced nature of newsrooms.

This study, therefore, examines how creators of public-facing data visualizations, especially in journalistic contexts, approach the task of designing for their audiences across media channels. Such practice-focused accounts are essential for identifying the challenges designers face, informing the development of tools, and providing a foundation for research communities in supporting practitioners working towards the common goal of inclusive data visualization. Hence, we take an exploratory approach, mapping current practices, assumptions, and challenges as described by practitioners rather than testing predefined hypotheses. Specifically, we address the following research questions:
\begin{enumerate} 
    \item \textbf{RQ-1:} Which factors and strategies do practitioners take into account when making design choices, and how do the envisioned target audiences and media channels influence those considerations? 
    \item \textbf{RQ-2:} To what extent, and in what ways, do practitioners evaluate visualizations (with their intended audience) in practice? 
\end{enumerate}

To address these questions, we conducted $21$ semi-structured interviews with data visualization practitioners from varied backgrounds and used thematic analysis to identify common practices and perspectives. We found a central tension: practitioners described audiences in broad terms and relied on platform specifics as proxies for user needs, yet stressed the need to design ``for the readers''. The main way interviewees differentiated audiences was by channel (including print, online, or social media), with these distinctions directly shaping design practices. Participants further emphasized conveying one clear message, telling a relatable story, and simplifying data as crucial multi-purpose strategies for reaching general audiences, even though tool support for these workflows remains scarce. Formal user testing was rare; instead, practitioners relied on peer feedback or social media metrics. The main contributions of this work are:

\begin{itemize}
    \item a qualitative account of how target audiences are understood and inform design,
    \item in-depth insights into how practitioners apply design strategies and considerations when creating for broad general audiences, 
    \item an exploration of current practitioner approaches to evaluating visualizations, and
    \item a set of design implications for HCI, identifying opportunities for rhetoric-aware and content-adaptive tools that support the visual communication of data to diverse audiences.
\end{itemize}

Together, these contributions provide a practice-oriented perspective on how public-facing data visualizations are targeted, designed, and evaluated, offering insights into the realities of journalistic visualization work.
\section{Related work} 
We situate our study in broader research investigating data visualization creation in practice for public-facing data visualizations geared toward lay audiences (Section~\ref{subsection:rb1}). We further examine work specifically focusing on how data visualizations are designed for different types of lay audiences and media channels (Section~\ref{subsection:rb2}) and, finally, how user evaluations are done in practice (Section~\ref{subsection:rb3}).

\subsection{Data visualization creation in practice}
\label{subsection:rb1}
In recent years, data visualization design practices have been studied from various angles, often by conducting surveys or interviews with practitioners to improve our understanding of their daily work habits. Many of such studies give valuable insights into very specific aspects of the design process, such as data exploration activities~\citep{alspaugh_futzing_2019}, client-designer negotiation~\citep{Lee-Robbins2024}, the role of inspiration and visualization examples~\citep{Bako2025,Baigelenov2025,bako_understanding_2023}, the integration of written language into the visualization design processes~\citep{Stokes2025}, team collaboration~\citep{zhang_how_2020}, or the ethical challenges of visualizing demographic data~\citep{Dhawka2025}. However, much work in the field focuses on opinions and viewpoints from data visualization practitioners working in non-journalistic settings, e.g., data analysts in industrial, academic, or regulatory environments~\citep{alspaugh_futzing_2019}, design students or professionals~\citep{bako_understanding_2023, bigelow_reflections_2014}, or data science employees at a large technology company~\citep{zhang_how_2020}.

Paul Parsons and colleagues have conducted a range of studies on visualization practice. In their 2021 interview study with 20 practitioners~\citep{parsons_understanding_2021}, they found that highly systematic approaches were uncommon, with participants instead \textit{``rely[ing] on situated forms of knowing and acting in which they draw from precedent and use methods and principles that are determined appropriate in the moment''}. Only one interviewee had a solely journalistic background and reported practices that differed from the rest. Other work by Parsons and colleagues likewise emphasizes the iterative and situated nature of design~\citep{parsons2025problem}, applies a philosophical framework of design judgment to examine how practitioners assess progress~\citep{parsons_design_2020}, and explores creativity and its role in practice~\citep{parsons_fixation_2021}. In addition, they surveyed designers about their knowledge and use of design methods (e.g., sketching) and principles (e.g., cognitive load)~\citep{parsons_what_2020}.

A recent study conducted in a more journalistic setting was the interview study by \citet{esteves_learned_2022}, who interviewed graphic reporters and data journalists about their academic backgrounds and daily tasks. They found that professionals rarely hold formal degrees in data visualization or data journalism but come from various educational backgrounds. The authors further report on the critical role of self-teaching and learning on the job in data visualization creation. \citet{engebretsen_visualization_2017} interviewed data visualization experts in $10$ news organizations in Norway, Denmark, and Sweden, and reflect on tensions in this field, ranging from complex, exploratory data visualizations vs. simpler, illustrative graphics to the role of ordinary journalists in data visualization production practices.

Extending this line of research, \citet{zhang_visualization_2022} interviewed COVID-19 dashboard creators from various organizations in the US to gain insights into their design process, as well as tensions, conflicts, and challenges that arose during that time. \citet{gregory_datajourneys_2024} traced the ``data journeys'' underlying climate change and COVID-19 visualizations in \textit{Scientific American}, documenting how data are sourced, transformed, and communicated in a popular science magazine. While providing rich insight into journalistic-like production processes, these settings differ from general newsrooms by targeting a more specialized, knowledge-oriented readership.  

Together, this body of work highlights diverse pathways into and within visualization work but sheds less light on the concrete strategies practitioners apply when designing for broad general audiences. Research specifically on journalistic settings is still limited, and largely fragmented, tending to focus on isolated aspects of design (e.g., the use of visualization examples), specific newsroom contexts (e.g., in specific countries), or particular visualization types or domains (e.g., COVID-19 dashboards). This type- and domain-specific focus makes it difficult to understand which design goals, considerations, and strategies practitioners employ in journalistic visualization work, where they typically handle a variety of topics and visualization types. Our study takes a complementary approach by examining cross-domain practices and challenges as they emerge in everyday journalistic work. 

Overall, these studies also reflect a broader tension observed in HCI: Parsons’s dual-gap model~\citep{parsons_understanding_2021} argues that generalization-oriented academic research often sits apart from the improvisational, judgment-driven, and context-dependent nature of professional visualization practice. More and more practice-oriented studies in the field describe how visualization work is often driven by tacit and situated knowledge, with theoretical principles adapted or abbreviated under real-world constraints. Yet, our theoretical and empirical understanding of practitioners' everyday design aims and strategies in journalistic settings is still developing, particularly in relation to how they create public-facing visualizations for broad and largely undefined audiences.

\subsection{Data visualization design for different audiences and channels} 
\label{subsection:rb2}
Beyond documenting workflows, another body of research has focused on design considerations themselves, examining how practitioners think about specific features of visualization design and, to a lesser extent, how these decisions intersect with distribution channels. Yet, channels are not just technical means of delivery: they shape how visualizations are perceived, constrain available formats, and influence the kinds of audiences that engage with them~\citep{schuster_being_2024,santos_instagrammable_2022}. Understanding design considerations without accounting for these channel-specific dynamics risks overlooking a central factor in how visualizations reach and resonate with public audiences.

\textbf{General design considerations.}
Many of the existing studies on design considerations focus on particular aspects, such as practitioners' perspectives on visual embellishments~\citep{parsons_data_2020} or on how to design thumbnails for data stories~\citep{kim_thumbnails_2019}. Previous work has further investigated the role of data stories~\citep{weber_data_2018, shao2024}, aesthetics and clarity~\citep{quispel_aesthetics_2018, kostelnick2020art, Kostelnick2008}, uncertainty~\citep{hullman_why_2020}, and data transparency~\citep{kennedy_data_2020} through a practitioners' perspective. Such studies typically isolate single aspects rather than documenting how practitioners navigate and apply design principles in everyday decision-making. Additionally, if at all, most of these analyses only briefly mention different channels of distribution, such as online, print, or social media. However, as news consumption increasingly shifts toward digital and mobile platforms, the role of distribution channels has become particularly important for how general audiences encounter and interpret visualizations~\citep{engebretsen_data_2020}.

\textbf{Design across media channels.}
Some studies on public-facing data visualization touch on distribution channels, often as part of how they define or reach their target audiences, for example, by using certain media channels such as news sources. Research foci reach from storytelling~\citep{segel_narrative_2010} to misconceptions~\citep{mena_reducing_2021} to transparency~\citep{kennedy_data_2020}. Notably, \citet{de_haan_when_2018} conducted an eye-tracking study to learn more about how news consumers use and appreciate data visualizations.  Contrary to many other works, they distinguish between different channels, such as print newspapers, e-newspapers on tablets, and news websites. While they found that news consumers use data visualization on all platforms, additional focus groups indicated \textit{``that visualizations online, and particularly on mobile devices, demand a new analysis on the form and function of visualizations to tell a news story''}~\citep{de_haan_when_2018}. Another study examining distribution channels is the 2022 analysis by \citet{yang_systematic_2022}, who analyzed COVID-19 information processing and argued that the target audience's beliefs about the communication channels affect how they process information. 

\textbf{Design for social media audiences.}
In particular, research interest in data visualizations on social media has been growing since the COVID-19 pandemic, with most studies either investigating the usage of COVID-19 data visualizations on social media channels~\citep[e.g.,][]{rotolo_coordinated_2022, li_communicating_2021,lee_viral_2021, trajkova_exploring_2020} or the effects and perceptions of these graphics on social media users~\citep[e.g.,][]{umaroh_pretesting_2023,al-nuwaiser_effect_2022,lee_perceptions_2022}. Beyond COVID-19, user engagement with infographics (in part as opposed to other forms of content) has been studied with outcomes that underline their importance~\citep[e.g.,][]{lochner_audience_2021,kunze_infographics_2021,kauer_public_2021}. In a small study interviewing $6$ Instagram users, \citet{burrows_sharing_2023} analyzed how participants engage with infographics and on which basis they would trust or share content. Despite social media gaining importance in research and practice, studies that focus on the design choices in social media data visualizations are still scarce. Examples of those include the investigations of \citet{spicer_creating_2022}, who share resources and advice on how researchers can utilize infographics and visual abstracts on social media to disseminate their research, as well as \citet{shu_what_2021}, who studied how data-GIFs are created in practice and how specific design choices in data-GIFs influence their understandability.

In summary, these studies highlight the growing relevance of distribution channels for how audiences encounter visualizations. Yet, most research either addresses channels only indirectly or focuses on narrow cases while isolating specific aspects of design rather than examining how practitioners systematically make and apply design decisions for general audiences across media contexts. From an HCI perspective, this gap is notable: theories of sensemaking and cognitive load emphasize that the interpretation of visualizations depends not only on design features but on the environments through which they are encountered. Channels shape how people attend to and process information, highlighting that platform specifics should be treated as an essential part of design considerations. It still remains unclear how practitioners integrate such considerations into concrete strategies for creating accessible visualizations for broad, undefined audiences, particularly as distribution channels evolve with social media.

\subsection{Evaluation and user testing in practice}
\label{subsection:rb3}
Another line of research relevant to our study concerns how visualizations are evaluated and the extent to which practitioners involve users (their envisioned audience) in this process. Evaluation is a crucial component of data visualization design, and the importance of evaluation practices has been acknowledged for many years~\citep{santos_evaluation_2014}. Prior work has shown that producers’ intentions do not always align with audience interpretations, and untested design strategies may risk misunderstanding or reduced trust~\citep{knoll_gulf_2025,schuster_being_2024,ridley_evaluating_2020}. User testing as well as approaches to directly involving users in the design process, such as participatory design~\citep{janicke_participatory_2020}, can help ensure that visualizations are understandable and effective for their intended publics. However, evaluation practices for public-facing data visualizations have received comparatively little attention in research. \citet{errey_evaluating_2023} conducted an interview study with $12$ data visualization practitioners, mostly in English-speaking countries. Their findings show that practitioners in their sample commonly approached evaluation informally, relying on their prior experience and colleagues to assess their work. Only about half of their study participants consulted end-users; social media channels like Twitter were essential evaluation tools for those who did. Overall, direct user engagement, user testing, and other forms of assessment remain under-explored in public-facing data visualization, with only limited insights available on journalistic contexts. 

In summary, these studies suggest that systematic evaluation plays a limited role in public-facing visualization practice. HCI literature on the gulf of evaluation, mental models, and sensemaking, along with recent empirical findings, highlight that alignment between designers’ intended messages and audience interpretation cannot be taken for granted.
Similarly, Human-Centered and participatory design principles emphasize involving users to ensure that visualizations meet the needs of their intended audiences. Yet empirical studies suggest that practitioners’ evaluation practices are largely informal, leaving only limited insight into how those in journalistic settings assess whether their visualizations work for broad, undefined audiences.

\setlength{\parskip}{1em} 
\setlength{\parindent}{0pt}

\textbf{Summary:} Existing research on public-facing data visualization has provided valuable insights, but largely concentrates on non-journalistic settings, a strict process-focus, or isolated aspects of design work. Meanwhile, the increasing use of visualizations in news media, particularly on digital platforms and social media, underscores the need to better understand how they are adapted for broad and diverse audiences. Yet, we still know relatively little about i) which design strategies and deliberations practitioners draw on when designing for different audiences and channels, and ii) how they evaluate their work in practice on the basis of their audience conceptions. Our study addresses this gap by examining how practitioners in journalistic spaces navigate audience- and channel-specific design choices and how they conceptualize the challenges of designing for broad and undefined audiences.

\section{Methodology}
We conducted semi-structured interviews with $21$ producers of public-facing data visualizations for general audiences. In this paper, we use the terms `general audiences' or `the general public' to refer to people who regularly or occasionally consume news media through communication channels, such as online, print, or social media. These persons may or may not have expertise in a given topic and may or may not perceive themselves as data (visualization) literate. This study was carried out as part of a project that has undergone ethical screening according to the guidelines of our institution and has been determined to be low-risk.

\subsection{Participants}

We used purposive and snowball sampling to recruit interview partners whom we contacted via e-mail. We sought diversity in job roles, institutional affiliations, and professional backgrounds. The key selection criterion was that participants regularly created data visualizations; i.e., visualization production was a core component of the participants' professional roles and not an occasional or supplementary task. The work of practitioners in our sample primarily focused on public-facing, communicative data visualizations for general audiences, rather than exploratory visualizations to aid data analysis in professional contexts. Some participants reported creating data visualizations, for example, for scientific purposes or companies; however, all participants also produced visualizations for general audiences. Although social media work was not a selection criterion, $19$ participants regularly created visualizations for social platforms or stated that their institution had a dedicated social media team, which would create data visualizations.

We selected participants based on their job roles and affiliated institutions, including news outlets, data visualization agencies, or other relevant organizations such as a statistical office or a science communication organization, aiming at general public data communication. In our participant sample, we also included freelancers and people who run social media channels, with a focus on data visualizations. While our focus is on journalistic settings, our sample includes not solely news outlets, as newsrooms increasingly rely on external specialists and agencies to produce visual content. Participants worked across a broad range of journalistic formats (e.g., explanatory pieces, breaking news graphics, scrollytelling projects, social media visuals). They were generalist visualization practitioners rather than domain specialists, typically producing graphics for multiple topical areas. We only contacted individuals who fit our criteria and interviewed all who agreed to participate. We concluded recruitment upon reaching theoretical saturation, where subsequent interviews confirmed existing themes without yielding novel insights into our core research questions~\citep{ladonna_beyond_2021}.

\begin{table}
\caption{Interview participants' IDs, institutions/clients (social media presence is mentioned only for dedicated data visualization accounts), job roles, and media channels they create data visualizations (vis) for.}
\begin{adjustwidth}{0cm}{}
    \footnotesize
    \centering
    \bgroup
    \def\arraystretch{1.2}
    \begin{tabular}{p{0.12\textwidth}p{0.35\textwidth}p{0.19\textwidth}p{0.043\textwidth}p{0.043\textwidth}p{0.043\textwidth}p{0.043\textwidth}} 
        \multirow{3}{*}{\parbox{1.7cm}{\textbf{ID}}} & 
        \multirow{3}{*}{\parbox{6.2cm}{\textbf{Participants' institutions and/or clients}}} & 
        \multirow{3}{*}{\parbox{1.7cm}{\textbf{Job roles}}} & 
        \multicolumn{4}{c}{\textbf{Used media channels}} \\
         & & &
        \multirow{2}{*}{\parbox{1.7cm}{{\scriptsize Online}}} &
        \multirow{2}{*}{\parbox{1.7cm}{{\scriptsize Print}}} &
        \multirow{2}{*}{\parbox{1.7cm}{{\scriptsize Social \newline Media}}} &
        \multirow{2}{*}{\parbox{1.7cm}{{\scriptsize Other}}} \\
         & & & & &  & \\
        \hline
        News-Orga-1 &
        Large national news agency offering vis services &
        Vis leadership role &
        \multicolumn{1}{c}{X} & \multicolumn{1}{c}{X} & & \\
        News-Orga-2 &
        Large national news magazine and website &
        Vis leadership role &
        \multicolumn{1}{c}{X} & \multicolumn{1}{c}{X} & \multicolumn{1}{c}{X} & \\
        News-Orga-3 &
        Large national daily newspaper and website &
        Vis leadership role &
        \multicolumn{1}{c}{X} & \multicolumn{1}{c}{X} & \multicolumn{1}{c}{X} & \\
        News-Orga-4 &
        Large national news website and TV channel &
        Data journalist &
        \multicolumn{1}{c}{X} & \multicolumn{1}{c}{X} & \multicolumn{1}{c}{X} & \multicolumn{1}{c}{X} \\ 
        News-Orga-5 &
        Large national daily newspaper and website &
        Infographic designer &
        \multicolumn{1}{c}{X} & \multicolumn{1}{c}{X} & \multicolumn{1}{c}{X} & \\
        \hline
        Graphics-Orga-1 &
        Agency for news media, institutions, corporate clients &
        Graphic designer, illustrator &
        \multicolumn{1}{c}{X} & \multicolumn{1}{c}{X} & \multicolumn{1}{c}{X} & \\
        Graphics-Orga-2 &
        Agency for news media, corporate clients &
        Founder of the organization &
        \multicolumn{1}{c}{X} & \multicolumn{1}{c}{X} & \multicolumn{1}{c}{X} & \\
        Graphics-Orga-3 &
        Agency for news media, corporate clients; social media &
        Founder of the organization &
        \multicolumn{1}{c}{X} & \multicolumn{1}{c}{X} & \multicolumn{1}{c}{X} & \\
        Graphics-Orga-4 &
        Agency for news media, corporate, scientific clients &
        Lead developer &
        \multicolumn{1}{c}{X} & \multicolumn{1}{c}{X} & \multicolumn{1}{c}{X} & \\
        Graphics-Orga-5 &
        Consultancy for corporate clients; social media &
        Founder of the organization &
        \multicolumn{1}{c}{X} & \multicolumn{1}{c}{X} & \multicolumn{1}{c}{X} & \\
        \hline
        Freelancer-1 &
        Freelancer for scientific clients; social media &
        Data vis designer, illustrator &
        \multicolumn{1}{c}{X} & \multicolumn{1}{c}{X} & \multicolumn{1}{c}{X} & \\
        Freelancer-2 &
        Freelancer for journalistic projects, books &
        Information designer &
        \multicolumn{1}{c}{X} & \multicolumn{1}{c}{X} & \multicolumn{1}{c}{X} & \\
        Freelancer-3 &
        Freelancer for news outlets, companies &
        Information designer &
        \multicolumn{1}{c}{X} & \multicolumn{1}{c}{X} & \multicolumn{1}{c}{X} & \\
        Freelancer-4 &
        Freelancer for news outlets, journalistic projects &
        Data journalist &
        \multicolumn{1}{c}{X} & \multicolumn{1}{c}{X} & \multicolumn{1}{c}{X} & \\
        Freelancer-5 &
        Freelancer for scientific clients, companies, books &
        Data vis designer &
        \multicolumn{1}{c}{X} & \multicolumn{1}{c}{X} & \multicolumn{1}{c}{X} & \\
        \hline
        Social-Media-1 &
        Private data visualization TikTok account &
        Biostatistician &
         &  & \multicolumn{1}{c}{X} & \\
        Social-Media-2 &
        Private data visualization Instagram account &
        Statistician, data analyst &
         &  & \multicolumn{1}{c}{X} & \\
        Vis-Magazine &
        Data visualization magazine &
        Graphic designer &
        \multicolumn{1}{c}{X} & \multicolumn{1}{c}{X} & \multicolumn{1}{c}{X} & \\
        Sci-Comm-Orga &
        Non-profit science communication organization &
        Leadership role &
         &  &  & \multicolumn{1}{c}{X} \\
        Statistical-Office &
        Statistical office &
        Communication officer &
        \multicolumn{1}{c}{X} & \multicolumn{1}{c}{X} & \multicolumn{1}{c}{X} & \\
        Vis-Company &
        Company offering a visualization tool and services &
        Founder of the organization &
        \multicolumn{1}{c}{X} &  & \multicolumn{1}{c}{X} & \multicolumn{1}{c}{X}\\
    \end{tabular}
    \vspace{0.2cm}
    \egroup
    \label{tab:participants-jobroles}
    \vspace{-5mm}
    \end{adjustwidth}
\end{table}

\autoref{tab:participants-jobroles} shows an overview of participants' institutions/clients, their job roles, and the media channels for which they create data visualizations. To protect the identity of our participants, we aggregated self-reported job roles at a higher level of abstraction. Participants were assigned an ID indicating their affiliation type (news organizations, graphics organizations, freelancers, social media, and others). Most practitioners had years of experience in the field, with half having 5+ years in their current role and some of them having worked for multiple organizations. Eight interviewees were either in leadership positions in their respective departments or had founded their own companies. Most participants were between $26$ and $35$ ($9$ participants) or between $36$ and $45$ ($6$ participants) years old. Participants were based in $10$ countries, with most working in Germany ($n=7$), Austria ($n=4$), and the USA ($n=3$). One participant each worked in Australia, Chile, Italy, Luxembourg, Sweden, Switzerland, and the UK. $19$ participants had an academic education, including $4$ with a PhD. Areas of education included natural sciences ($5$), social sciences ($5$), arts ($3$), formal sciences ($2$), and applied sciences ($1$). Three people had an academic background in journalistic fields, and a further $3$ in visual/graphic communication/design. One person had a non-academic education as a graphic designer.

\subsection{Semi-structured interviews}
The interviews were conducted between December $2022$ and June $2023$. Each interview lasted about one hour and was conducted in person or via Zoom, depending on participant preference and location. Interviews were conducted in English or German. The interview guideline followed a structured format covering key areas of interest. However, questions were adapted according to participants' backgrounds and the conversation flow. Interview partners were asked to bring a data visualization representative of their work to support discussion of their creation process and design decisions. This approach helped elicit detailed, experience-based responses. The discussion revolved around the following themes: (i) Introduction and consent; (ii) Demographics; (iii) Design process and practices; (iv) Design considerations, evaluation practices, and user insights; (v) Data visualization creation during the COVID-19 pandemic. COVID-19-related questions explored how the pandemic may have influenced workflows or design considerations. The complete interview schedule is provided in the supplementary material (S1); exemplar questions include:
\begin{itemize}
    \item Based on the example visualization you sent, what does a typical data visualization creation process look like?
    \item Imagine a new colleague starting tomorrow. What are the most important things to tell them?
    \item What role does your target audience play in creating a visualization? 
    \item How do you ensure that a data visualization is understandable?
    \item How do you get feedback on your data visualization designs, or how do you evaluate them?
\end{itemize}

After the interviews were recorded and transcribed, we analyzed the transcripts in the original language; quotes from interviews conducted in German were translated into English for this paper. We followed Braun and Clarke's framework for reflexive thematic analysis~\citep{braun_using_2006}, and used a combination of inductive and deductive coding~\citep{robson_real_2017}. While the first author led the analysis, key decisions and interpretive challenges were collaboratively discussed within the research team. To begin with, each transcript was read in full at least twice, with initial observations noted during the process. The first seven interviews were then coded using the qualitative analysis software Atlas.ti, forming the basis for an initial set of codes. Codes were clustered into tentative categories and candidate themes; the evolving codebook was reviewed and refined through discussions with three senior researchers at several stages. Codes and candidate themes were expanded, merged, and restructured as the remaining interviews were analyzed. We concluded data collection when we observed that the codebook had stabilized, that is, when patterns began to recur and no new insights were emerging with respect to our research questions~\citep{ladonna_beyond_2021}. Following this, codes for all $21$ interviews were reviewed using the final version of the codebook to ensure consistency and revisit earlier interpretations. The codes were clustered into $12$ themes relevant to the research questions, which are structured across four domains of interest. These domains, themes, exemplar subthemes and codes, as well as illustrative quotes can be found in the supplementary material (S2).

\section{Findings} \label{section:findings}
The following sections present four main areas of insight from our analysis: the evolving visibility and institutional role of data visualization in journalism (Section~\ref{subsection:role}), practitioners' notions of their target audience (Section~\ref{subsection:audience}, RQ-1), strategies for designing for broad general audiences (Section~\ref{subsection:strategies}, RQ-1), and the evaluation approaches participants described (Section~\ref{subsection:testing}, RQ-2). Quotes are provided with a participant ID indicative of the participant's group affiliation (e.g., News-Orga-1 for a person working for a news organization) as described in \autoref{tab:participants-jobroles}. 

\subsection{Journalistic data visualization in transition: new demands, old structures}
\label{subsection:role}
Interviewees acknowledged the importance of compelling data visualizations for public communication, particularly for topics with a high societal impact, such as COVID-19 or climate change, where much of the argumentation is based on or informed by data: \textit{``If we add more climate change data into a repository, it's not going to do anything. But if we can more effectively communicate to leaders, individuals, stakeholders, that's how we start to move the needle forward.''} (Vis-Company). Visualization and dashboard development during the COVID-19 pandemic was described as extremely fast-paced and chaotic, as readers were expecting daily up-to-date graphics: \textit{``The only time in my career before COVID when we had data at the moment where there was also interest in data, people were looking for it, was election nights. With COVID, every day was kind of election day. [...] Until then, we spent weeks and months on projects and then struggled to get attention.''} (Freelancer-4). One biomedical researcher described how the urgency of the pandemic led him to open a data visualization social media channel to share results directly with the public, contrasting this immediacy with the slow pace of journal publications. The pandemic was also said to have brought lasting changes, raising the public profile of data visualization and prompting several practitioners to launch social media accounts to meet growing interest: \textit{``The pandemic changed people's point of view of the importance of data and data visualization.''} (Vis-Company).

Participants also noted shifting news consumption habits, with most organizations now producing content ``mobile or online first.'' Traditional channels are still used, but digital platforms and specific outlets with a stronger focus on data visualizations are gaining importance: \textit{``The weekend edition gets more and more important because reading newspapers is something that [fewer] people do than a couple of years ago. [...] More young people look at these weekend editions and digital projects that are a little bit more visual.''} (News-Orga-5). While the significance of social media varied across participants' work, most acknowledged its growing importance for data communication. Some described how they actively consider social media throughout the visualization design process: \textit{``We always try to think [about] social media during the whole design process as well. So sometimes you have, during the design process, already the feeling: Okay, these could be cool Instagram cards in the end because you have different aspects.''} (News-Orga-2).

These shifts in how data visualizations are produced and consumed are also reflected in organizational structures. Visualization departments, once seen as service units producing on-demand graphics, are increasingly integrated into journalistic workflows: \textit{``Five years ago or so, people said: `I have this data. Can you give me a line chart of that?' [...] Now, we move more and more to a department [...] also involved in the content.''} (News-Orga-2). Despite this shift, several challenges persist. The continued underappreciation of visualization work in many organizations is reflected in limited budgets and the late involvement of visualization designers in editorial processes (often only after the article text has already been written). Interviewees spoke of these changes with great enthusiasm, while some also expressed clear anger at the shortcomings that persist. Across settings, workflows were described as highly iterative, emphasizing flexibility over fixed procedures, which was often named as necessary to meet diverse project and client demands. In newsrooms, time-sensitive publication cycles and required editorial sign-offs add further constraints. Tool choices reflected this same need for flexibility: interviewees most frequently mentioned using Adobe Illustrator, Datawrapper, or R, illustrating a trade-off between visual design tools and the more specialized functionality of dedicated visualization tools. Finally, professional pathways into visualization are still varied and informal, with many practitioners entering from diverse backgrounds, often lacking formal training, and learning on the job and developing tacit knowledge instead.
\setlength{\parskip}{1em} 
\setlength{\parindent}{0pt}

\textbf{Summary:} Together, these developments highlight a field in transition: data visualizations are gaining visibility and importance across platforms and organizations, but (newsroom) structures and professional norms have not fully kept pace. Practitioners continue to adapt to new demands, often without the benefit of formal design training or strong institutional support.

\subsection{Who is the audience?}
\label{subsection:audience}

All participants emphasized that understanding one's audience is crucial for creating effective data visualizations. However, in practice, readerships were often only vaguely defined. When asked about their target audience, a freelancer remarked: \textit{``That's not always clear. [...] If you ask the client, they're not even aware of it.''} (Freelancer-5).  Most relied on broad and ambiguous definitions, such as the ``general public.'' Only a few described their typical audience in more concrete terms, such as age, gender, education level, or language. An example of a more elaborate distinction came from one participant, who categorized users and charts on two dimensions, namely passive vs. active and informing vs. entertaining, to guide decisions about using simple charts or more illustrative/interactive formats. The most prominent high-level distinctions typically focused on practical parameters such as expertise, age, or preferred consumption channels.

A common contrast was between expert and non-expert users, with most assuming a non-expert general audience by default. While experts were seen as more tolerant of complex or information-rich formats, practitioners emphasized that accuracy was equally important across both groups. However, balancing understandability with accuracy was described as a central challenge in tailoring visualizations to audience expertise. Age was also frequently mentioned: older audiences were associated with traditional channels and design styles (e.g., white backgrounds), while younger audiences were described as more visually literate, quicker content consumers, and more inclined to rely on social media platforms such as Instagram or TikTok, often as their primary news source: \textit{``They want to see just the post on Instagram and that one caption at the bottom that kind of summarizes the story. While the older generation will sit and read the whole article.''} (Graphics-Orga-3). 

Most notably, all interviewees observed that audience requirements and reactions vary strongly across media channels. Print and desktop formats were described as more flexible, allowing for layered information, complex layouts, and detailed annotations. However, online readers were seen as largely mobile-first, requiring visuals to remain legible on small screens. Social media poses the most distinct demands, with graphics having to function as stand-alone visuals. Even across social media platforms, user expectations and behaviors were said to differ: the same visualization could go viral on one channel but fail on another, shaped both by algorithms and the distinct audience preferences on each platform. These differences make it difficult to serve all channels with a single visualization, often requiring entirely separate designs.
\setlength{\parskip}{1em} 
\setlength{\parindent}{0pt}

\textbf{Summary:} Practitioners consistently stressed the importance of knowing their audience, yet definitions were often vague, with most defaulting to the broad ``general public'' Only occasionally did participants distinguish more concretely, for example, by expertise, age, or channel, which then shaped design requirements.

\subsection{Tailoring without a clear target: designing for the ``general public''}
\label{subsection:strategies}
Four core aims for designing data visualizations for general audiences emerged from the interview analysis: \textit{catching attention} (Section \ref{subsubsection:attention}), \textit{supporting comprehension} (Section \ref{subsubsection:comprehension}), \textit{fostering credibility} (Section \ref{subsubsection:credibility}), and \textit{fostering a connection} (Section \ref{subsubsection:emotion}). Strategies to achieve these aims are often interrelated and sometimes in tension. Participants further highlighted multi-purpose strategies inherently relevant to multiple aims (Section \ref{subsubsection:crossstrategies}). \autoref{tab:strategies} summarizes the most frequently mentioned aims and strategies that practitioners rely on in everyday practice; \autoref{tab:tips} lists shared lower-level guidelines. While some participants mentioned certain domains (e.g., climate or COVID-19) as more sensitive, they did not describe domain-specific design strategies. Accordingly, our analysis focuses on cross-cutting audience-related practices rather than domain-specific differences. While a few interviewees referred to formal design principles or user testing, most drew on experience accumulated through years of hands-on work designing and publishing visualizations across different outlets and client contexts.

\begin{table}[t]
\caption{Practitioner strategies for designing data visualizations for broad audiences, organized by four aims (attention, comprehension, credibility, connection) plus multi-purpose practices. The theme blocks synthesize recurring approaches reported by interviewees, with comprehension often prioritized over other aims.}
\setlength{\tabcolsep}{6pt}              
\renewcommand{\arraystretch}{1.25}       
\centering
\small 

\begin{tabularx}{\linewidth}{
  >{\RaggedRight\arraybackslash}p{.27\linewidth}
  >{\RaggedRight\arraybackslash}p{.69\linewidth}
}
\rowcolor{Butter}
\multicolumn{2}{c}{\rule{0pt}{3.0ex}\normalsize\textbf{Making people look: Strategies to catch attention}}\vspace{1mm}\\
\rowcolor{white}
\textbf{Make it beautiful} & Use aesthetically pleasing designs and balanced layouts/palettes to draw the eye. \\
\textbf{Make it novel and engaging} & Add surprising details or uncommon forms sparingly to spark curiosity. \\
\textbf{Use formats/topics that attract} & Favor chart types and topics that reliably pull audiences in, such as rankings. \\
\end{tabularx}

\vspace{3mm}

\begin{tabularx}{\linewidth}{
  >{\RaggedRight\arraybackslash}p{.27\linewidth}
  >{\RaggedRight\arraybackslash}p{.69\linewidth}
}
\rowcolor{SoftBlue}
\multicolumn{2}{c}{\rule{0pt}{3.0ex}\normalsize\textbf{Helping people understand: Strategies to support comprehension}}\vspace{1mm}\\
\rowcolor{white}
\textbf{Ensure clarity} & Use unambiguous, readable titles, labels, and annotations. \\
\textbf{Guide the reader} & Use hierarchy, annotations, and stepwise structure to direct attention. \\
\textbf{Use familiar chart types} & Stick to familiar chart types like bar/line charts, or maps that audiences recognize. \\
\textbf{Explain abstract concepts} & Clarify abstract concepts using text, visual aids, or metaphors. \\
\end{tabularx}

\vspace{3mm}

\begin{tabularx}{\linewidth}{
  >{\RaggedRight\arraybackslash}p{.27\linewidth}
  >{\RaggedRight\arraybackslash}p{.69\linewidth}
}
\rowcolor{SoftGreen}
\multicolumn{2}{c}{\rule{0pt}{3.0ex}\normalsize\textbf{Building trust: Strategies to foster credibility}}\vspace{1mm}\\
\rowcolor{white}
\textbf{Stay truthful to the data} & Rely on credible sources, apply fact-checking, and ensure statistical accuracy. \\
\textbf{Be transparent} & Show sources directly on graphics, and correct mistakes openly. \\
\textbf{Explain limitations} & Communicate data limitations clearly and explain uncertainties. \\
\textbf{Favor proven styles} & Use standard encodings and meaningful color, avoiding misleading embellishments. \\
\end{tabularx}

\vspace{3mm}

\begin{tabularx}{\linewidth}{
  >{\RaggedRight\arraybackslash}p{.27\linewidth}
  >{\RaggedRight\arraybackslash}p{.69\linewidth}
}
\rowcolor{Lav}
\multicolumn{2}{c}{\rule{0pt}{3.0ex}\normalsize\textbf{Evoking emotion: Strategies to foster connection}}\vspace{1mm}\\
\rowcolor{white}
\textbf{Use color deliberately} & Apply palettes that evoke the right tone, avoiding either dullness or alarmism. \\
\textbf{Treat sensitive topics carefully} & Use neutral styles or language, and cultural sensitivity when visualizing serious issues. \\
\textbf{Establish a connection} & Connect to readers by showing local or personal aspects of data. \\
\end{tabularx}

\vspace{3mm}

\begin{tabularx}{\linewidth}{
  >{\RaggedRight\arraybackslash}p{.27\linewidth}
  >{\RaggedRight\arraybackslash}p{.69\linewidth}
}
\rowcolor{LightGray}
\multicolumn{2}{c}{\rule{0pt}{3.0ex}\normalsize\textbf{Strategies serving multiple purposes}}\vspace{1mm}\\
\rowcolor{white}
\textbf{Convey one clear message} & Ensure that every visualization communicates one central takeaway. \\
\textbf{Tell a relatable story} & Frame data in ways that connect to people’s experiences and context. \\
\textbf{Make it simple} & Reduce complexity by focusing on what audiences can reasonably grasp. \\
\textbf{Adapt design to the platform} & Adjust visuals to fit the medium and audience expectations, particularly for social media. \\
\end{tabularx}
\vspace{3mm}
\label{tab:strategies}
\end{table}

\subsubsection{Making people look: Strategies to catch attention}
\label{subsubsection:attention}
Capturing attention was described as a necessary first step in engaging general audiences with data visualizations. Participants often emphasized aesthetics as crucial for drawing viewers in and making charts appealing. However, they stressed that beauty alone is not enough and that clarity must never be compromised. Views on embellishments such as icons, illustrations, or creative details varied. Still, most found them useful for enhancing appeal and sparking engagement: \textit{``It's just an image [depicting the chart topic] I put on top. But it was mentioned so often by people that this was something why they stopped. [...] Following the rules, this would not be allowed at all. This is not adding anything to this story; it is not supporting the data. But I think it works exactly like this: Exciting people and kind of briefing them quickly on the topic.''} (Freelancer-5). Strong engagement was also said to be linked to topical relevance and storytelling techniques that invite personal identification. Country comparisons were frequently cited as reliable attention grabbers, as readers instinctively look for their own country. Maps, in particular, were often reported to outperform other chart types in engagement metrics.

\subsubsection{Helping people understand: strategies to support comprehension.}
\label{subsubsection:comprehension}
Designing visualizations that are easy to understand for audiences with varying levels of visual literacy was a central concern. Besides ensuring clarity in annotations, labels, and titles, guiding readers through a chart was emphasized as an effective simplification practice, for example, by using highlights or stepwise builds: \textit{``What we really like to do is highlighting. So, not showing all the data at once, and here you go. But finding a focus on a certain number on a certain line in a multi-line chart, for example, and just explaining that and putting everything else more in the background.''} (News-Orga-2). Some also described interactive designs that reveal detail progressively, allowing users to start with a simplified view and access more complex data as needed. There was broad agreement that familiar chart types support comprehension best. Bar, line, pie charts, and maps were most frequently used, while more elaborate designs were often considered less understandable and sometimes even prompted requests for simpler alternatives: \textit{``I'm quite active on social media, getting feedback like: `Well, this should be a bar chart'.''} (Freelancer-5). Some practitioners avoided pie charts due to interpretability concerns, and scatter plots were often seen as too demanding for general audiences. Participants also pointed to concepts that many readers struggle with, including percentages, probabilities, large numbers, logarithmic scales, and double axes: \textit{``If it's something that you cannot really see in a way, but you have to have a mental concept first before understanding it, then it gets really difficult.''} (Sci-Comm-Orga). In these cases, textual explanations were considered essential.

\subsubsection{Building trust: strategies to foster credibility.}
\label{subsubsection:credibility}
Trust was closely tied to accuracy, reliability, and transparency. Practitioners stressed the importance of using credible sources, fact-checking, and showing data sources directly on the graphic. Some also emphasized being open about inconsistencies or limitations rather than hiding them. Views diverged on whether to include uncertainty: several participants advised against it, arguing that most audiences struggle with such concepts and that visualizations should stand on their own. Others, however, considered uncertainty integral to honest reporting, especially in areas like climate change: \textit{``Simplifying it would be like not showing very important information. So this is something where I'd say: Okay, information comes first, and also kind of the truth to the information, and not so much the user experience.''} (Freelancer-2). Standard design choices, such as shaded intervals or dotted lines, were seen as relatively intuitive ways to visualize uncertainty, even though accompanying text was still regarded as crucial.

\subsubsection{Evoking emotion: strategies to foster connection.}
\label{subsubsection:emotion}
Evoking emotion was described as particularly important in sensitive contexts such as the COVID-19 pandemic. Practitioners highlighted the role of deliberate color choices and restrained design, cautioning that overly polished aesthetics can create distance. Instead, they sometimes used more approachable styles, including hand-drawn elements, to foster approachability and connection. Several emphasized revealing the human stories behind numbers to counteract ``psychic numbing,'' where large counts of deaths or cases risk losing emotional impact. Humanizing strategies included showing specific locations on maps or integrating photos, quotes, or short texts to bring people into the narrative. However, participants noted that icons or illustrations can backfire in tragic contexts, underscoring the need for sensitivity.

\subsubsection{Strategies serving multiple purposes.}
\label{subsubsection:crossstrategies}
While the strategies discussed above were usually linked to a specific aim, some approaches were repeatedly described as serving multiple purposes, helping practitioners catch attention, support comprehension, foster credibility, and evoke emotion at the same time.

\textbf{Convey one clear message.} Without being prompted, most participants stressed the importance of defining a clear message to make a visualization understandable for broad audiences. Several emphasized that the message should guide the entire design process: \textit{``I always ask the client, what is that one [piece of] information that you want to get through? [...] There is not only the information but also some kind of a message.''} (Graphics-Orga-1). A descriptive title was seen as a particularly effective way to convey the key takeaway. While titles are often used for chart metadata (e.g., units), participants argued they should instead communicate the main point of the chart: \textit{``The title should basically be the thing the reader learns from that chart. So it will not be what is shown, but what you learn.''} (Freelancer-4).

\textbf{Make it simple.} Simplification was often described as essential for making complex topics graspable, but also as a core strategy for other communicative aims such as attracting attention or fostering trust. Some interviewees called `learning to simplify' the single most important advice for newcomers: \textit{``I think the most important thing is [...] don't overcomplicate anything. I mean, it's quite nice to get carried away and do lots of nice fancy things, but actually the simpler the visualization, the better.''} (News-Orga-4). Practitioners stressed the effortful nature of simplification, with reducing while striving for accuracy being named as one of the hardest challenges on the job. While some cautioned that simplification can go too far, others argued that simplicity should be prioritized, even if nuance is reduced.

\textbf{Tell a relatable story.} Many participants described storytelling as a powerful tool to engage readers and convey complex information. Rather than focusing solely on data presentation, they highlighted the value of crafting a coherent narrative, or as one interviewee put it: \textit{``How to make charts and diagrams, it's not that rocket science. But if [...] you can't tell a story, and you can't focus on what's important, then you can't do infographics.''} (News-Orga-1). Storytelling was also seen as a discipline that benefits newsroom workflows more broadly in the sense that data visualization can force editorial teams to sharpen their focus, as it requires journalists to distill the data into one clear and condensed storyline.

\textbf{Adapt design to the platform.} 
Practitioners emphasized that social media requires especially bold design choices: Larger fonts, brighter colors, cheeky language, and striking visuals condensed to one clear message were described as necessary to get people's attention, even if these choices felt exaggerated and the \textit{``graphic designer's heart [might] bleed''} (Vis-Magazine). Some also pointed to short-form video, where charts build themselves up with music and pacing, as particularly effective. These formats reflect the split-second decisions of social media users: \textit{``Social media, in all honesty, you're gone in a jiffy if you need anything more than three brain cells. [...] The point of social media is just: Information in your face. [...] `Here, hello, read this, watch this, leave a like'.''} (Vis-Magazine).
 
\setlength{\parskip}{1em} 
\setlength{\parindent}{0pt}

\textbf{Summary:} Practitioners emphasized four main aims when designing for broad audiences: capturing attention, supporting comprehension, fostering credibility, and fostering connection. To achieve these goals, they employed approaches such as aesthetics, transparency, and sensitivity, complemented by multi-purpose practices including conveying a clear message, sharing relatable stories, simplifying content, and tailoring their approach to social media platforms.

\begin{table}[!htbp]
\caption{Low-level guidelines for designing data visualizations for broad audiences, as reported by practitioners.}
\setlength{\tabcolsep}{5pt}
\renewcommand{\arraystretch}{1.12}
\centering
\small

\begin{tabularx}{\linewidth}{>{\RaggedRight\arraybackslash}p{.17\linewidth} >{\RaggedRight\arraybackslash}X}
\toprule
\textbf{Theme} & \textbf{Low-level guidelines practitioners named in the context of designing for a ``general audience''} \\
\specialrule{.1pt}{2pt}{2pt}

\textbf{Message and Framing} &
  $\bullet$ Define the main takeaway before designing.\newline
  $\bullet$ State the key message in the title; describe what is shown in the subtitle. \\
\specialrule{.1pt}{2pt}{2pt}

\textbf{Labels and Annotations} &
  $\bullet$ Label key peaks, thresholds, or changes directly on the data.\newline
  $\bullet$ Minimize reliance on legends; label elements directly when possible.\newline
  $\bullet$ Print units or symbols (\%, €, k) on the chart, not only in captions.\newline
  $\bullet$ State the unit once (e.g., in the title) rather than repeating it on every axis label. \\
\specialrule{.1pt}{2pt}{2pt}

\textbf{Color and Typography} &
  $\bullet$ Keep entity colors consistent across spreads.\newline
  $\bullet$ Use high-contrast, clearly distinguishable hues.\newline
  $\bullet$ Test palettes for color-blind accessibility.\newline
  $\bullet$ Align color with tone (e.g., avoid alarm red in neutral contexts).\newline
  $\bullet$ Keep typography consistent (fixed fonts, distinct print vs.\ digital). \\
\specialrule{.1pt}{2pt}{2pt}

\textbf{Chart Types and Encodings} &
  $\bullet$ Default to bar and line charts for general audiences.\newline
  $\bullet$ Avoid pie charts, scatterplots, dual axes, and log scales unless necessary; explain them when used.\newline
  $\bullet$ Use maps for pattern, not precision; normalize values (e.g., per capita) and pair with bars for detail.\newline
  $\bullet$ Use small multiples to prevent overplotting.\newline
  $\bullet$ Replace big numbers with unit blocks/dots (e.g., ``1 dot = 10 people'') to convey magnitude.\newline
  $\bullet$ Introduce encodings step by step rather than all at once. \\
\specialrule{.1pt}{2pt}{2pt}

\textbf{Layout} &
  $\bullet$ Establish a clear focus point, with supporting elements in hierarchy.\newline
  $\bullet$ Use white space and grids to structure content and group elements.\newline
  $\bullet$ Alternate simple charts with occasional complex ones to sustain attention.\newline
  $\bullet$ In print, guide attention along reading flow (upper left $\rightarrow$ lower right).\newline
  $\bullet$ In digital, surface graphics early in the scroll; avoid hiding behind generic photos.\newline
  $\bullet$ Ensure every element earns its place; avoid embellishments without purpose. \\
\specialrule{.1pt}{2pt}{2pt}

\textbf{Text} &
  $\bullet$ Use plain language and avoid jargon.\newline
  $\bullet$ Use text to highlight the takeaway and explain abstract concepts. \\
\specialrule{.1pt}{2pt}{2pt}

\textbf{Interaction} &
  $\bullet$ Provide simple defaults and hide advanced options in drill-down views.\newline
  $\bullet$ Keep interactive controls minimal to avoid overwhelming first-time users.\newline
  $\bullet$ Avoid exploratory UIs for general audiences; guide them through the story instead.\newline
  $\bullet$ Use concise on-chart instructions for interactions (e.g., ``Enter your postcode…''). \\
\specialrule{.1pt}{2pt}{2pt}

\textbf{Credibility and Transparency} &
  $\bullet$ Verify data carefully; a single error can undermine trust.\newline
  $\bullet$ Prefer official or primary sources; avoid secondary aggregators unless fully traceable.\newline
  $\bullet$ Print the data source directly on the graphic.\newline
  $\bullet$ Show inconsistencies or caveats with visible warnings; do not publish if a dataset is unreliable.\newline
  $\bullet$ Correct mistakes publicly by taking down and republishing.\newline
  $\bullet$ Explain uncertainty with clear visuals (e.g., shaded confidence bands, dotted forecasts) and text. \\
\specialrule{.1pt}{2pt}{2pt}

\textbf{Tone, Sensitivity, and Connection} &
  $\bullet$ Match icons or illustrations to the seriousness of the topic; avoid playfulness for tragic issues.\newline
  $\bullet$ Keep visuals clean and captions neutral for sensitive topics.\newline
  $\bullet$ Bring data closer to lived experience with photos, illustrations, or icons.\newline
  $\bullet$ Use metaphors to make abstract risks tangible (e.g., probabilities, survival rates).\newline
  $\bullet$ Consider approachable styles for sensitive topics (e.g., hand-drawn).\newline
  $\bullet$ Reveal the people behind the data through narratives, photos, or quotes. \\
\specialrule{.1pt}{2pt}{2pt}

\textbf{Social Media} &
  $\bullet$ Adapt font sizes and colors for different devices; keep language concise.\newline
  $\bullet$ Add illustrations or icons to increase interest.\newline
  $\bullet$ Consider short-form video for fast-paced platforms.\newline
  $\bullet$ Use bold design choices that prompt users to pause while scrolling.\newline
  $\bullet$ Reduce content to one clear main message. \\
\specialrule{.1pt}{2pt}{2pt}

\end{tabularx}

\label{tab:tips}
\end{table}

\subsection{The gap between valued feedback and formal evaluation} 
\label{subsection:testing}

Feedback on visualization design in general was highly valued among participants and even mentioned by some as the most important success factor for data visualization creation. Several emphasized that true evaluation requires audience feedback, yet formal evaluation methods were rarely used by the practitioners in our sample. Instead, they often reported relying on their own judgment to assess a visualization's clarity and appeal. At the same time, this reliance on personal experience or intuition was not universally accepted, and some interviewees strongly criticized the assumption that designers can reliably assess audience needs without external input: \textit{``This needs knowledge about what people know or not. And you yourself are not the measurement. [...] But I need another intelligence. This is knowing the limitation of my audience.''} (Graphics-Orga-2).

While most organizations in our sample did not conduct user testing, a few exceptions stood out. The statistical office described running usability panels with moderated remote sessions, short surveys, and task-based feedback (user needs, expectations, navigation, pain points), which sometimes lead to actionable changes. The national news agency reported occasional efforts, such as in-the-wild tests where passers-by (e.g., in McDonald's or Starbucks) are asked for quick like/dislike reactions. Apart from these cases, user testing was largely absent, partially due to organizational disinterest: \textit{``There isn't this wish for [user testing], so we don't test it on anyone but ourselves, and so far it has worked that way.''} (Vis-Magazine). Others pointed to tight deadlines, limited budgets, and rapid publication cycles that left little room for structured pre-release evaluation, particularly in news agencies. Participants also noted that other industries, such as product design or tech, were more likely to invest in structured feedback loops: \textit{``Industries, other than media out there, are much better at doing this. They really measure; they make qualitative interviews like you do now. They ask their customers, `Do you like this and that more?'''} (Graphics-Orga-2).

In the absence of formal testing, internal feedback was considered essential. Drafts were routinely shared with peers, often conducted as quick ``hallway tests'' seeking ``fresh eyes'' from someone not involved in the project. Freelancers, on the other hand, often relied on iterative feedback loops with clients. Senior editors frequently acted as a proxy audience: \textit{``The editor-in-chief is the representative of the audience, he represents the users, and says: `Okay, nobody can understand this. Nobody's interested in this. Nobody likes this. This graphic is ugly.' ''} (News-Orga-1).

Participants reported on collecting performance metrics after publication or release, although many described these as insufficiently informative. For visualizations embedded in articles, they noted it is nearly impossible to interpret common measures such as click-through rate, dwell time, or scroll depth independently of the article's topic, timing, or placement on the website. As a result, analytics were seen as indicating (circumstantial) attention but not comprehension. Despite their similar nature, practitioners considered social media feedback more valuable, with views, likes, shares, and comments offering some sense of reach. Direct responses, such as critical Twitter threads, Reddit discussions, or private messages, were widely valued as an important source of post-release critique. For larger projects, some organizations also hold internal reflection meetings a few weeks after publication to discuss lessons learned and document takeaways.
\setlength{\parskip}{1em} 
\setlength{\parindent}{0pt}

\textbf{Summary:} While practitioners consistently valued feedback, formal evaluation practices, such as user testing, were largely absent in their workflows. Instead, designers employed informal methods, including internal feedback and post-release audience reactions, particularly on social media. This reflects a gap between the recognized importance of (audience) testing and the practical means to implement it, which limits opportunities for systematic learning.
\section{Discussion}
In this study, we interviewed $21$ practitioners who create data visualizations for journalistic purposes to investigate how they make design decisions with respect to their envisioned target audiences (RQ-1) and how they assess their designs in practice (RQ-2). This focus on situated design decision-making and real-world evaluation practices contributes to ongoing discussions in HCI research on sensemaking, visual data communication, inclusivity, and practitioner-centered tool design in light of broad and diverse audiences. 

Understanding how practitioners design data visualizations requires attending to their organizational context, as recognition and resourcing within these organizations strongly influence what is possible in practice. Our findings revealed a tension: audience demand for data visualizations has grown, particularly during the COVID-19 pandemic~\citep{Arafat2023,zhang_visualization_2022}, yet in some media organizations, visualization work remains underappreciated. However, practitioners described a gradual shift toward greater recognition and involvement in content production, indicating progress from the circumstances described in a 2012 study by~\citet{weber_data_2012}. This shift from service provider to creator of narrative, multi-channel data stories extends the scope of their responsibilities. This evolution requires workflows and tools that extend beyond chart production and support practitioners in addressing broader communication, editorial, and audience-related challenges.

In this section, we contextualize key findings (summarized in \autoref{tab:discussion_summary}) in light of prior research, outline implications for designing tools and methods that better align with practitioners' workflows, and identify opportunities for future work that supports audience-aware visualization design. Specifically, we examine how the realities surfaced in the interviews point to new approaches in five main areas: audience definitions (Subsection~\ref{discussion-1-audience}), platform adaptations (Subsection~\ref{discussion-2-platforms}), design strategies for attention (Subsection~\ref{discussion-3-attention}), rhetorical guidelines (Subsection~\ref{discussion-4-rhetoric}), and evaluation methods (Subsection~\ref{discussion-5-evaluation}).

\begin{table*}[ht]
\caption{Summary of main insights, relation to prior work, and opportunities for tool development.}
\setlength{\tabcolsep}{4pt}
\renewcommand{\arraystretch}{1.3} 
\footnotesize
\begin{tabularx}{\linewidth}{
  >{\raggedright\arraybackslash}p{0.11\linewidth} 
  >{\raggedright\arraybackslash}p{0.34\linewidth}  
  >{\raggedright\arraybackslash}X                  
}
\toprule
\textbf{Theme} & \textbf{Key interview insights} & \textbf{Relation to prior work \& opportunities for tool development} \\
\midrule

\textbf{Defining the Audience} \newline \textit{(Section~\ref{discussion-1-audience})} &
\textbullet\ Target audience is typically vaguely defined as ``the general public'' \newline
\textbullet\ Practitioners rely on tacit user knowledge \newline
\textbullet\ Simplification as default strategy for non\textendash expert users, assuming low attention and data (vis) literacy &
Builds on~\citep{burns_who_2023,bottinger_reflections_2020} with a practical perspective; vague audience notions are consistent with journalism studies~\citep{dewerth2013audience}; connects to HCI discussions of inclusive design and equity \newline
\textit{Opportunity:} Minimal practical audience definitions and ``view as'' modes to simulate diverse literacy levels/constraints within the tool
\\
\midrule

\textbf{Platforms as Proxies} \newline \textit{(Section~\ref{discussion-2-platforms})} &
\textbullet\ Platforms as proxies for user needs/preferences \newline
\textbullet\ Greater integration of vis work in editorial processes and shift to mobile-first/multi-channel \newline
\textbullet\ Manual (time-intensive) rebuilding of charts for different formats, sometimes by extra teams &
Validates/extends~\citep{de_haan_when_2018,weber_data_2012} by emphasizing the changing conditions and need for platform adaptations \newline
\textit{Opportunity:} Content-adaptive systems that support multi-format export for different channels beyond simple responsive layouts, and platform ``preview'' modes that make platform constraints explicit in the design process
\\
\midrule

\textbf{Designing for Attention} \newline \textit{(Section~\ref{discussion-3-attention})} &
\textbullet\ User attention in journalistic contexts is scarce \newline
\textbullet\ ``Stopping the scroll'' is a major design aim \newline
\textbullet\ A clear chart message acts as a cognitive handle, e.g., in the form of an ``explanatory'' title &
Connects to~\citep{wanzer_role_2021, weber_data_2012}; reframes effectiveness for public-facing visualization as contingent on capturing initial attention \newline
\textit{Opportunity:} Message-first workflows, LLM consistency checks, and saliency prediction
\\
\midrule

\textbf{Rhetorical Aims} \newline \textit{(Section~\ref{discussion-4-rhetoric})} &
\textbullet\ Practitioner aims beyond attention span: comprehension, credibility, connection \newline
\textbullet\ Storytelling as a common multi-purpose strategy \newline
\textbullet\ Tensions between simplicity and accuracy &
Contextualizes perspectives from~\citep{prantl_untangling_2025,Kostelnick2008,Hullman2011} by connecting practitioners' aims to rhetorical theory \newline
\textit{Opportunity:} Rhetoric-aware tools supporting semantic styling and smart simplification
\\
\midrule

\textbf{Evaluation Gap} \newline \textit{(Section~\ref{discussion-5-evaluation})} &
\textbullet\ Formal user testing is often infeasible \newline
\textbullet\ Reliance on intuition and peer feedback \newline
\textbullet\ Social signals (e.g., comments, likes) are proxies for success but lack depth &
Confirms~\citep{errey_evaluating_2023}; contrasts with empirical tools~\citep{choi_vislab_2023} highlighting a persistent gulf of evaluation, despite feedback being regarded as crucial \newline
\textit{Opportunity:} Automated in-tool checks and micro-feedback workflows to operationalize participatory evaluation in newsrooms
\\

\bottomrule
\end{tabularx}
\label{tab:discussion_summary}
\end{table*}

\subsection{From ``general public'' to minimal practical audience definitions} \label{discussion-1-audience} 

It is undisputed in academic and practitioner communities that ``knowing your audience'' is essential for creating customized content and designing effective data visualizations~\citep[e.g.,][]{drucker_communicating_2018,hakone_proact_2017}. This premise becomes more challenging to address when the intended audience is broad or undefined, and when charts must appeal to a wide range of readers~\citep{bottinger_reflections_2020}, for example, including both experts and non-experts in any given field. Most interviewees in our sample referred to their audience as ``the general public.'' This resonates with prior observations in journalism studies on newsrooms, as~\citet{dewerth2013audience} wrote: 
\begin{quote}
\textit{Although journalistic images of the audience may be `incomplete,' they do exist and powerfully help shape the work of journalists in producing journalistic texts. [...] It is also misleading to argue that reporters do not `know' their audience simply because they are not keyed into demographic information about them. Much of what journalists know about their audience  [...] is imbedded in their work routines and is not easily expressed.}
\end{quote}

Similarly, in our interviews, participants rarely provided detailed accounts of their audience demographics. However, they reported a wealth of tacit knowledge about what works for their readers. A common default strategy for designing effective public-facing data visualizations was simplification, assuming a baseline of low attention and low data (visualization) literacy. This was described as both strategic and inevitable, aligning with prior work~\citep[e.g.,][]{engebretsen_what_2023,schuster_being_2024}. At the same time, simplification also raises concerns: what is reduced or left out can shape interpretation~\citep{dignazio_data_2023}, sometimes at the cost of nuance or accuracy. A key instance of this dilemma is uncertainty: practitioners often exclude uncertainty in their charts to prevent overwhelming readers, but visualizing it can increase transparency and trust~\citep[e.g.,][]{hullman_pursuit_2019,corell_ethical_2019,yang_uncertainty_2024}. Yet, trust is also closely related to the clarity and simplicity of a chart; it reflects a highly personal judgement that is dependent on specific user characteristics, further emphasizing ``the importance of designers understanding their target audience''~\citep{mckinley_trustworthy_2025}.

While much visualization research has pointed to the need (and the challenge) of defining lay audiences~\citep{burns_who_2023}, our findings highlight a different set of questions: To what extent \textit{can} practitioners define such audiences, and what is feasible from a design and resource standpoint in journalistic settings? Our findings suggest that precise audience definitions are often neither possible nor practical. Newsroom readership spans a wide range of people; reaching as many of them as possible is not only an idealistic goal but also a key driver of profit~\citep{edgerly2024speaking}. Visualizations must, therefore, resonate with readers of varying skill sets and backgrounds and the common practice of ``designing for everyone'' risks becoming ``designing for no one'' in particular. This raises broader questions about inclusive design and links our study to HCI debates on equity and fairness; specifically, whether visualizations are appropriate across different literacy levels, abilities, or sociocultural backgrounds~\citep{knoll_gulf_2025}. 


Our findings underscore the need for audience-aware design strategies that can accommodate diverse readers without relying on rigid or unrealistic audience definitions. This suggests opportunities for lightweight ways of characterizing audiences, such as simple audience-modeling tools and adaptive or platform-specific design templates that fit into fast-paced newsroom workflows. Future research could explore minimal practical audience definitions, such as simplified personas specific enough to guide design while remaining feasible in newsroom environments (e.g., ``the distracted scroller,'' ``the colorblind reader,'' ``the low-data-literacy reader''). Tool builders could implement these as ``view as'' modes, allowing designers to simulate different audience types directly within authoring tools. Such in-tool checks could support adaptive or personalized designs that tailor complexity according to different literacy levels or preferences~\citep[e.g.,][]{shin_drillboards_2025,gupta2016visualization}. Smart simplification modes, in which users can adjust detail through interaction, may further ease the tension between simplification and accuracy. Where such adaptability is not feasible, it remains crucial to study how diverse audiences perceive simplified visualizations and where they draw the line between helpful clarity and misleading reduction.

\subsection{The platform as audience proxy: opportunities for adaptive design} 
\label{discussion-2-platforms} 

The challenge of accurately conceptualizing ``the general public'' led our interviewees to design their visualizations not so much for \textit{who} the audience is, but for \textit{where}, i.e., on which platform, the audience is. This means they rely on platform specifics rather than user demographics as a primary influencing factor for their design decisions. Considerations include typical user demographics (e.g., Facebook's comparatively older vs. TikTok's younger-leaning user base) and preferences (e.g., high-level summaries on Instagram vs. deeper engagement on a website), as well as technical requirements associated with offline, online, or specific social media platforms.

Although prior work highlights the need for cross-platform adaptation~\citep{de_haan_when_2018}, there remains a lack of empirical design guidance, particularly for social media. Existing advice mostly comes from industry sources such as Tableau~\citep{tableau_tips_2022} or Flourish~\citep{flourish_guide_2023}, offering broad recommendations such as ``know your goal and audience''~\citep{linkedin_how_nodate}. Our findings suggest that compared to other platforms, social media requires bold visual design choices, reduced complexity, and sometimes animation.

In light of the current tool landscape, practitioners in our sample often had to manually rebuild charts for their social media platforms. Future research could investigate how adaptive tools, potentially supported by AI, might help practitioners generate multiple chart versions for different platforms. Unlike traditional responsive design, which simply resizes the layout, these tools should be content-adaptive, adjusting not only the layout but also visual complexity, structure, and narrative emphasis. This would allow designers to define a dataset and auto-generate distinct content pieces, such as a web version (granular data, interactive elements, textual explanation) and a social media version (aggregated data, bolder design, static format). Such tools could also include platform emulators (e.g., previewing a chart in a mock Instagram UI) to help designers evaluate content and the socio-indexical ``vibe''~\citep{morgenstern_visualization_2025} within an actual feed.

However, these systems must align with journalistic values to avoid becoming ``technically impressive but [...] practically useless''~\citep{xiao_it_2025}. As AI makes adaptive content more feasible, the challenge shifts from manual creation to guiding and constraining these systems so they produce outputs that are not only audience-aware but also editorially responsible and contextually appropriate. While our interviewees did not report using automation or AI in their workflows, recent work highlights generative approaches~\citep{Thomson_Generative_2024,chen2024how}, and some outlets have announced plans to employ AI for visualization production~\citep{kasper2023deutsche}. This suggests an opportunity to explore how such tools can be developed and integrated into journalistic practice and other forms of visual data communication.

\subsection{Designing the hook: capturing attention in a discretionary context}
\label{discussion-3-attention} 

This reliance on platform specifics has further implications for visualization design: participants emphasized that stopping the user from scrolling is a necessary precursor to any effective communication. ``Stopping the scroll'' is an established concept in digital marketing; yet, visualization research has often viewed attention-seeking strategies with skepticism, prioritizing analytic depth over immediate appeal. However, our findings highlight a pragmatic reality: public-facing visualization operates within an attention economy. Unlike professional analysts who must use a tool, public audiences choose to engage. In this environment, the first user ``task'' is deciding whether to pause. This aligns with recent findings by~\citet{shi_double_2025}, who demonstrated that social media users filter visualizations based on an initial ``overall look'' before engaging further. This decision often depends on the visualization's ``vibe''~\citep{morgenstern_visualization_2025}, meaning whether or not the user perceives the design as something they identify with. We argue that designing for attention is not necessarily about ``clickbait'' but can be about ensuring that information is visible in contexts where the amount of content is abundant but cognitive resources are limited (e.g., when doom-scrolling through social media while on public transportation). At the same time, this sits in tension with well-founded critiques of the attention economy: attention-seeking strategies risk reproducing online behavioral mechanisms that are rightfully viewed as harmful. Our findings, therefore, highlight a broader dilemma: practitioners must make information noticeable without replicating exploitative dynamics of online attention systems.

To prevent users from scrolling past the visualization, a clear and concise message was the most frequently cited success factor. While prior work stresses the importance of such takeaways primarily for understandability~\citep{adar_communicative_2021,bateman_useful_2010,gammelgaard_ballantyne_images_2016,koesten_what_2023,Stokes2025}, our findings suggest that in the feed, the message serves a more immediate function: reducing the interaction cost. As news media charts are often perceived as too complex~\citep{knoll_gulf_2025,schuster_being_2024}, a clear message can act as a cognitive handle; i.e., users might be more likely to invest their time, as the perceived mental load and potentially actual decoding effort are reduced. To address this, some practitioners in our study argued that titles should explicitly convey the main message~\citep[e.g.,][]{kong_frames_2018,borkin_what_2013}, an approach supported by~\citet{wanzer_role_2021}, who found that informative titles reduce mental effort and increase perceived aesthetics. 

Our findings suggest that designing for attention becomes a necessary precursor to other, more established design premises for public-facing data visualizations. Consequently, beyond asking ``How do laypeople read charts?'' HCI research should also try to answer ``How do charts capture attention in the feed?''. This suggests that design requirements should also focus on the specific rhetorical aspect of guiding readers' cognitive focus and future studies should account for realistic consumption contexts, including what motivates users to pause or interact. 

To formalize the strategies observed in our interviews, we propose that visualization tools should support a ``message-first'' workflow for dedicated communicative purposes (distinct from analytical workflows where titles are often generated last). For instance, systems could prompt designers to draft the takeaway message before selecting a chart type. LLMs could then be used to perform automated ``consistency checks,'' i.e., flagging instances where the title's or the chart's claim contradicts the underlying data trend or intended message. To gain a better understanding of which design features are most effective in promoting user engagement, tools could also integrate AI-driven saliency maps to predict the ``visual hook'' during the design phase. This would allow creators to verify whether specific design choices potentially fulfill their purpose of catching the user's attention without requiring expensive eye-tracking studies. Drawing on practitioners' insights that personal relevance also drives engagement, tools could automatically suggest localized views for geospatial data.

\subsection{Navigating rhetorical tensions: towards rhetoric-aware tools}
\label{discussion-4-rhetoric} 

Beyond catching the audience's attention, practitioners described three core aims: supporting comprehension, fostering credibility, and fostering connection. These aims can be understood as forms of rhetoric, where communication is adapted to its purpose and audience~\citep{kostelnick2003shaping, Kostelnick2008}.
Rhetorical theory reminds us that effective communication engages not only logic but also credibility/trust (\textit{Ethos}) and emotion (\textit{Pathos})~\citep{campbell2019feeling}.
While visualization research has traditionally prioritized clarity and efficiency (\textit{Logos})~\citep{tufte_visual_1983}, rhetorical theory suggests that truly effective messages strike a balance between these appeals~\citep{prantl_untangling_2025}. 

Our findings show that practitioners actively employ \textit{Ethos} (e.g., through transparency about data sources) and \textit{Pathos} (e.g., by humanizing data points). This aligns with recent findings that practitioners often employ value-laden design approaches~\citep{Dhawka2025} and confirms the stance that data visualizations are not neutral artifacts, but rhetorical devices that can frame data and alter audience perception and interpretation~\citep{prantl_untangling_2025, Hullman2011}.

Practitioners emphasized storytelling as a technique they use to balance these rhetorical aims; that is, they employ a narrative structure in their visualizations to support comprehension, credibility, and connection~\citep[cf.,][]{segel_narrative_2010, Shukla2025,Milesi2025}. However, these aims come with trade-offs, and navigating them can be a challenging task. For instance, storytelling strategies often aim to ``humanize'' data to foster empathy~\citep{boy_showing_2017}; yet, this often conflicts with the need for perceived neutrality, as minimal chart design is traditionally viewed as objective and factual~\citep{kennedy_work_2016}. Recent work, on the other hand, warns that this ``scientific'' aesthetic is not universally effective: for some audiences, overly polished visuals can actually signal inauthenticity or manipulation~\citep{morgenstern_visualization_2025}. Practitioners currently rely on intuition to resolve these tensions, lacking formal guidance on how to style a visualization that is both engaging and trustworthy. Our findings, therefore, point to opportunities for more explicit guidance and tooling that help practitioners reason about the rhetorical consequences of their design choices, including how transparency cues or humanizing elements shape audience perception. They also suggest the need for lightweight methods that surface rhetorical trade-offs early in the design process, making it easier to balance appeals to logic, trust, and emotion in fast-paced production environments.

Regarding tools, practitioners described a trade-off between dedicated data platforms (e.g., Datawrapper, Tableau) and free-form design software (e.g., Illustrator). The former were often criticized as inflexible, echoing recent findings that built-in templates can constrain creativity~\citep{Baigelenov2025}. We propose ``rhetoric-aware systems,'' that is, visualization tools that move beyond tone-neutral templates and incorporate semantic styling functionality. Building on work regarding affective learning objectives~\citep{lee-robbins_affective_2022} and emotional engagement~\citep{Kennedy2018}, a designer can, for example, specify a rhetorical goal (e.g., urgent, engaging) and the system suggests visual encodings (e.g., color palettes, layout) that align with that tone~\citep{prantl_untangling_2025}. To support \textit{Pathos}, tools could integrate anthropographic features, such as pictographs and icons, which allow designers to represent data through relatable human figures.

\subsection{Closing the evaluation gap with lightweight methods}
\label{discussion-5-evaluation}

Practitioners in our sample relied primarily on tacit knowledge and professional intuition, a pattern also noted in previous work~\citep{errey_evaluating_2023,parsons_understanding_2021}. While this reliance reflects years of experience, it highlights a tension: practitioners recognized that their own expertise often does not align with audience perspectives. This concern is well-founded; recent work by~\citet{knoll_gulf_2025} demonstrated a ``gulf of interpretation'' where audience takeaways frequently diverge from producers' intent. Similarly, readers often form strong social judgments about the social provenance of a chart that the designer never intended~\citep{morgenstern_visualization_2025}. However, constraints of time and resources typically make formal user testing infeasible. While recent systems, such as VisLab~\citep{choi_vislab_2023}, have successfully streamlined empirical experiments, our findings suggest that even simplified experiments are often too slow in the fast-paced news cycle. This highlights a broader challenge: developing methods and tools that help practitioners externalize, interrogate, and refine their tacit assumptions about audiences within time-limited settings of newsroom production.

Consequently, the question arises: what does ``good enough'' evaluation look like in this context? Since we cannot change the speed of news, we argue that we must change the tools we use. Future work could explore how to move evaluation inside the authoring environment. While generic LLMs like ChatGPT can offer broad design feedback, they often lack the nuanced contextual understanding required for specific visualization problems~\citep{kim_how_2025}. To catch errors without leaving the workflow, we propose automated in-tool checks. These could flag specific issues, such as accessibility barriers or complexity thresholds (e.g., text density), prior to publication. At the same time, it is essential to recognize that this approach carries the risk of potentially marginalizing important user perspectives that are not incorporated into the evaluation logic. 

Beyond automation, tools should formalize the informal feedback strategies practitioners already use. Participants noted that social signals, such as likes and comments, are used as proxies for success, but lack depth. Conversely, asking a peer, e.g., through hallway testing, yields qualitative insights but is often unstructured. Future tools could formalize these behaviors into feasible ``micro-feedback workflows'': For in-house feedback, this could mean private preview links to send out to peers for annotations or to editors for sign-offs; for public audiences, this could mean embedding lightweight feedback mechanisms directly on news websites, such as minimal voting buttons (e.g., ``Helpful'' vs. ``Confusing'') and commentary functions specific to the visualization rather than the article. This would effectively transform the consumption context into a continuous, participatory evaluation environment, capturing reader signals in the wild without the friction of a comprehensive survey.

\subsection{Limitations}

Our findings must be interpreted within the limitations of the study's setup. Our primary sampling goal was to cover a breadth of professional backgrounds within our study focus; consequently, factors such as cultural diversity were secondary to professional role selection. While our sample includes practitioners based in ten countries across Europe, North and South America, and Australia, it remains centered on the German-speaking region, particularly Germany and Austria. Specific regional circumstances, such as local media systems, may therefore shape the findings. The study primarily reflects perspectives from Western contexts, where norms and infrastructure may differ significantly from other contexts. Furthermore, our analysis did not systematically distinguish between journalistic content domains (e.g., news, sports, finance, culture), reflecting the fact that the interviewed practitioners operated across various content areas and rarely cited domain-specific needs as design drivers. Although participants came from varied backgrounds, the distribution of occupational settings may not fully reflect the diversity of public-facing data visualization production. As interviewees were recruited via professional networks and snowball sampling, there may be a bias toward well-connected, visible practitioners. Similarly, interviewees may be more invested in the topic or more reflective than the average creator, potentially holding stronger views~\citep[e.g.,][]{parsons_understanding_2021,moore_statistics_nodate}.

As with other interview studies, our data reflect practitioners' self-reported practices, which may be shaped by ideals, retrospective rationalizations, or assumptions about ``good'' practice rather than offering direct insights into everyday workflows. Decisions described as strategic (e.g., color choice) may have been intuitive or default choices in the moment. Crucially, we analyzed the practitioner's beliefs and imaginings of the audience, rather than the audience itself. Reported design strategies are therefore based on professional intuition and may not fully align with actual user behavior. Finally, the semi-structured interview design may have prompted participants to focus on the predefined themes, and the interpretive nature of thematic analysis reflects the researchers' analytical lens. With 21 interviews, the study prioritizes depth of understanding over representativeness. The findings should therefore be interpreted as exploratory and analytically transferable, rather than statistically generalizable to all practitioners.

\section{Conclusion}  
Public-facing data visualizations play a vital role in public communication and democratic discourse, yet creators face the challenge of designing for an elusive, undefined audience under strict time constraints. Our study shows how practitioners navigate this by treating platform specifics as proxies for user needs. We identify four central design aims: attention, comprehension, credibility, and connection. We summarize the strategies practitioners use to balance these goals, such as simplification and narrative framing, while noting that they often rely on intuition and informal peer feedback rather than formal testing. To address these realities, we argue that visualization tools must evolve. We propose rhetoric-aware systems that support content adaptation for different media channels and integrate automated heuristics for quality control. Ultimately, tool design must reflect that public visualization is not merely about data transmission, but about capturing attention in discretionary consumption contexts. By aligning tool design with the high-velocity nature of newsroom production, the HCI community can help bridge the gap between academic rigor and the practical realities of public data communication.

\begin{acks}
We thank our interview partners for their collaboration, support, and time. This work has been funded by the Vienna Science and Technology Fund (WWTF) [10.47379/ICT20065].
\end{acks}

\bibliographystyle{ACM-Reference-Format}
\bibliography{bibliography}


\end{document}